# Mechanism for self-formation of periodic structures on a plastic polymer surface using a nanosecond and femtosecond laser pulses

*Behzad Mansouri*[1,*], *P.Parvin*[1], *Kelvin J. A. Ooi*[2]

[1]*Department of Physics, Amirkabir University of Technology, Hafez Ave, Tehran, Iran P.O. Box: 15875-4413*
[2]*Engineering Product Development, Singapore University of Technology and Design, 20 Dover Drive, Singapore 138682*
[*]*Corresponding author.* behzad.mansouri@hotmail.com

The high UV laser dose at 193 nm induces micro grooves on poly allyl diglycol carbonate PADC (CR39) at normal irradiation. The spatial period exhibits to be nearly invariant for azimuth and polar angles indicating a loose dependence on the incident angles but the LIPSS (Laser-induced periodic surface structures) are always parallel to the P polarization component of the incident beam. The most common approach to explain LIPSS formation is related to the Sipe theory which does not account for all the observed phenomena especially LIPSS with periodicity larger than the laser wavelength. In fact the LIPSS is a multi parameter mechanism based on surface rippling, acoustic modulation and laser ablation etc. In experiment with CR-39 polymer, laser irradiation produces a very tiny melting layer containing mixture of monomers due to depolymerization on the surface and it seems capillary wave is responsible for the grooves formation.

## I. INTRODUCTION

Laser-induced periodic surface structures (LIPSS) which have been studied since the 1960s [1] appear on dielectric, semiconductor, polymer and metal surfaces exposed to single or multiple short and ultra short laser pulses [1,2]. The most common LIPSS, as the surface ripples can be produced on metals [2], semiconductors [3], dielectrics [4] and polymers [5].When created with linearly polarized laser radiation at normal incidence, these ripples show a periodicity close to the laser wavelength, the ripples having these properties are usually attributed to the low spatial frequency LIPSS (LSFL). Furthermore, a new kind of ripples has been recently observed. Applying picoseconds or femtosecond laser shots, the ripples with a periodicity significantly smaller than the laser wavelength, referred to as high spatial frequency LIPSS (HSFL). LIPSS with periodicities larger than the laser wavelength have also been observed similar to those referred in [9]. In fact, irradiation of CR39 sample with excimer nanosecond ArF laser .These LIPSS, usually looks like as "grooves" [6]. Although, the exact physics of their formation remains unexplained we investigate the formation mechanism from different points of view here.

## II. ELECTROMAGNETIC APPROACH

It is generally accepted that LSFL formation is driven by the interaction of electromagnetic waves with the rough surface of materials [3]. Considering surface roughness that may shift the dispersion curve leads to coupling between the surface plasmon wave and the incident laser pulse. However, above mechanism can be dominant after several few shots to induce some roughness on our initial sample, the experimental observation demonstrated that grooves as shown in Fig.1can take place even after first shots.

The LIPSS mechanism for metals has been also explained with excitation of the surface plasmon polariton (SPP) [7, 8]. The femtosecond pulse plasma lengths scale is very short and we consider surface plasma. SPPs are electromagnetic excitations propagating at the interface between a dielectric and an electrically conductive material. Indeed, the ripple formation links to SPP in the case of semiconductor material like silicon using femtosecond pulses. Although, the generation of surface plasmon is due to the presence of free electrons at the interface of two materials that implies one of material should be a metal, it is suposed that under ultra-short laser exposure the optical properties of dielectrics change to be conductive. Regarding this consideration SPP excitation can be achieved. The parametric process of photon→photon+plasmon is referred to as stimulated Raman scattering where an incident photon decays into Plasmon and a scattered photon at lower frequency. Sakabe et al [8] showed that the ripples periodicity is shorter than the laser wavelength under femtosecond irradiation. The dispersion relation of surface plasmon indicates that the periodicity is ranging 0.85-0.5 of laser wavelength. While SPPs can explain a periodicity $\Lambda$ of LSFLs smaller than the laser wavelength ($\Lambda \leq \lambda$), it is unclear why $\Lambda$ depends on the fluence and decreases with the number of pulses applied. Only a few studies on the formation of LIPSSs with a periodicity larger than the laser wavelength ($\Lambda > \lambda$) as shown in Fig.1 are reported in the literature. In our setup the spatial period $\Lambda$ exhibits to be nearly invariant for azimuth $\varphi$ and polar $\theta$ angles indicating a loose dependence on the incident angles. However, the created lines are always parallel with the P polarization component of the incident beam. For nanosecond pulse the front phase of a laser pulse heats sample, which turns into the plasma subsequently. The plasma is heated by the tail of the pulse due to IB absorption (Inverse Bremsstrahlung) which expands at sonic speed. Consequently, the major part of the pulse makes more dense bulk plasma. Direct coupling laser pulse with plasma wave occurs when plasma frequency increases with the laser fluence as long as to match with laser frequency. Let assume the grooves formation in our experiment on CR39 sample occurs due to plasma wave, increasing the fluence will lead to increase plasma density, $n \propto F$ and as well as plasma frequency $\omega^2 = n(4\pi e^2)/m$, that makes decrease in wavelength and periodicity. Even in femtosecond case, surface electron density, $n'$ is proportional with $n/v$, $v$ is plasma velocity that directly correlate with $\sqrt{T}$, so $n' \propto n/v \propto F^{1/2}$ [8]. As one can see both case leads to the inverse relation between periodicity and laser fluence. This model does not completely match with the observation that illustrated in Fig 2 [9].

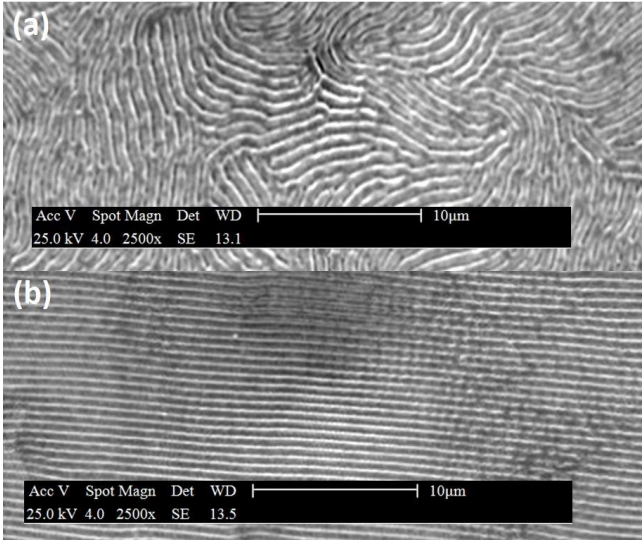

**Figure 1:** Laser-induced periodic surface structures on CR39 implied fringe pattern distribution. The fringe width or periodicity increase with increasing laser fluence within ranging 0.7- 0.8 $\mu m$ [9].

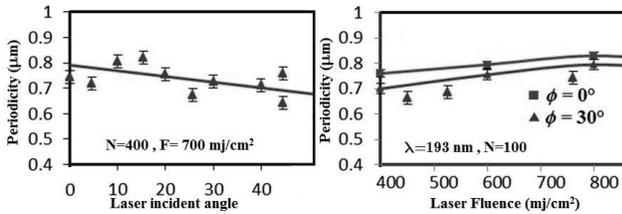

**Figure 2:** The groove periodicity versus incident fluence at $\varphi=0°$ (rectangles) and $\varphi=30°$ (triangles) incident angles and versus incident fluence [9].

## III. LASER INDUCED CAPILLARY WAVE

The exact theoretical description of all physical process is so extensive and complicated. In this paper we present concise analytical models. We consider capillary wave formation due to presence of surface tension in the molten layer on the surface due to depolymerization of polymer as a mechanism for self-formation of periodic fringe patterns at normal incidence.

ArF laser (LPX 220i, Lambda Phyzik) was employed with 15 ns duration and 1 Hz repetition rate to be spatially distributed over ~0.9cm × 2.5cm beam cross-section (CR39 sheets was cut in to 1.5mm × 10mm × 10mm), is focused on samples through a cylindrical $MgF_2$ lens (f = 20 cm) [9]. The laser profile is uniform in X direction and hat-top in Y direction which produces temperature gradient within the thickness of sample obviously but no temperature gradient across the thickness, so with this consideration, surface tension is constant and we can ignore Marangoni flow effect. Samples have been irradiated with various fluences, with multiple shot numbers and create melting layer on surface which become cool and solid before the next pulse comes. According to our calculation the order of magnitude of melting life time is about 1.6 μs and time duration between successive shots is 1s. For estimation spatial period length we start from dispersion relation for the capillary wave as below [10, 11]:

$$\omega = \sqrt{k(g + \frac{\alpha k^2}{\rho})\tanh(kh)} \qquad (1)$$

With rearranging based on wavelength we have:

$$\Lambda = \sqrt{\tfrac{1}{2}ght^2 + \frac{\sqrt{h}t\sqrt{16\pi^2\alpha + g^2ht^2\rho}}{2\sqrt{\rho}}} \qquad (2)$$

For small wavelength only the surface tension is dominant and we ignore gravity effect simplifying the equation into:

$$\Lambda = \sqrt[4]{\frac{\alpha h}{\rho}}\,(2\pi\tau)^{1/2}. \qquad (3)$$

Where $\Lambda$ is the wavelength of the capillary wave, α, h, ρ and τ are the surface tension coefficient, melt depth, mass density and period of wave respectively. Note that wave cannot exist more than life time of melting layer, so we can replace τ with $\tau_{melt.}$

CR-39 is a thermosetting plastic and cannot be melted in general. But when laser pulses interact with sample, a lot of processes due to photothermal, photochemical occurs and consequently leads to depolymerization of polymer into mixtures of monomers just on the surface of sample and finally change the optical and the thermophysical properties. However, the exact nature of the decomposed product is not well analyzed yet but the products of laser-induced decomposition are available in liquid. With considering thermogravimetric analysis of heat treated CR-39 polymer [12], melting is observed as small peak in the DTG plot of derivative weight against temperature.

Depending on type of laser pulse (short or ultra short) analyzing and modeling of laser ablation process and melting dynamic may be changed. In the case of femtosecond, because of shockwave presence, the mechanism would not be simple. The energy transport in ultrashort laser heating is two stages process. At first the absorption of the laser energy through photon-electron interactions within the ultrashort pulse duration and electrons reestablish the Fermi distribution in just a few femtoseconds. The second stage is the energy distribution to the lattice through electron-phonon interactions, on the order of tens of picoseconds. Two-temperature model is necessary in order to describe the exact thermal process. Despite we explained the fs laser interaction, our experimental data has focused on the ns laser shots. On the other hand, in the nanosecond case, the process is modeled by heat conduction, melting and plasma formation. For the estimation of melting thickness; first we consider temperature profile as below:

$$T_{melting} - T_{initial} = \frac{F}{\rho C \delta_P}, \qquad (4)$$

$T_{melt}$ is the working point temperature at which the sample can easily be formed with viscosity less than $10^3$ Pa.s. $\delta$ is optical penetration depth and F is absorbed laser energy fluence given by Beer–Lambert absorption law in term of incident fluence or fluence threshold and ablation depth:

$$F = AF_i \exp\left(-\frac{z}{\delta_P}\right) = AF_{th} \exp\left(-\frac{z - h_a}{\delta_P}\right), \qquad (5)$$

we get a formula for initial melting thickness as below:

$$h_M = \delta_P \ln\left(AF_{th} / (T_M - T_i)\rho C \delta_P\right), \qquad (6)$$

In order to estimate the surface absorptivity (A), we should note that however CR-39 is completely opaque in the UV range and in fact the reflectivity can be derived by the Fresnel equation which is about 4% in this case, but when sample is radiated with high fluence of laser pulses, reflectivity increases as the plasma density increases. Therefore with presence of the dence plasma, most of incident laser energy is reflected back from the surface (induced skin effect). In fact most of energy utilised to expand the plasma. We observe when the incident laser fluence is higher than threshold (applied fluence 700mj/cm$^2$ > $F_{th} \approx$ 225mj/cm$^2$), plasma with a critical density can be formed and a large portion of the energy is reflected from the sample.

For the estimation of the average melt time, we need to see the variation of temperature with time and place. According to our initial and boundary conditions, the solution of diffusion equation would be in terms of erfc as below:

$$\frac{\partial T}{\partial t} = D \frac{\partial^2 T}{\partial z^2}, \qquad (7)$$

$$T(z,0) = T_i + \frac{AF_i}{\rho C \delta_P} \exp\left(-\frac{z}{\delta_P}\right), \qquad (8)$$

Non-dimensionalized equation provides better insight into relative size of terms, so we have:

$$T' = \frac{\rho C \delta_P}{AF_i}(T - T_{initial}), \qquad (9)$$

$$z' = \frac{z}{\delta_P}, t' = \frac{D}{\delta_P^2} t, \qquad (10)$$

Finally we obtain a analytical solution for temperature distribution as below:

$$T'(z',t') = \frac{\int_0^\infty e^{-a}[e^{-\frac{(z'+a)^2}{4t'}} + e^{-\frac{(z'-a)^2}{4t'}}]da}{(4\pi t')^{1/2}}, \qquad (11)$$

$$T'(z',t') = \frac{e^{t'-z'}}{2}\left(\text{Erfc}\left[-\frac{-2t'+z'}{2\sqrt{t'}}\right] + e^{2z'}\text{Erfc}\left[\frac{2t'+z'}{2\sqrt{t'}}\right]\right). \qquad (12)$$

Now we can estimate both initial melt depth and average melt time. The average melt depth varies between 0.2 and 0.4µm and the average of melt lifetime varies between 0.8 µs and 1.6 µs. We used the thermophysical properties of decomposed products as mixtured of monomers on average in our calculation [13,14,15,16].

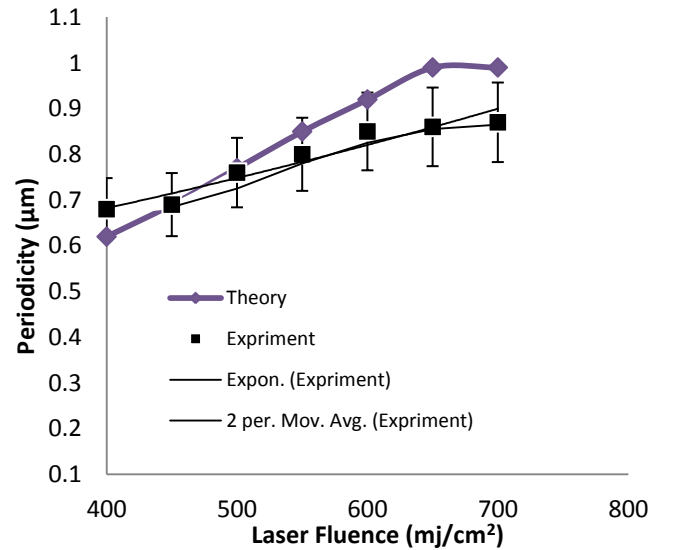

**Figure 3:** Comparison of the rather agreement between theory and experiment for periodicity versus incident laser fluence at normal incidence.

In summary, as shown in Fig.3, the increasing of periodicity with increasing laser fluence is consistent with capillary wave-driven formation. Increasing laser fluence leads to more deposited energy in the sample that gives rise to higher melting rate and make increase in liquid layer depth and consequently life time. Dispersion formula (3) exhibits that longer life time and liquid depth we have longer capillary wavelength and obviously longer periodicity. Also one can see the theoretical slope is higher than the experimental slope which indicates competitive mechanisms, plasma oscillation which is induced by laser shots (with its inverse relation between periodicity and the laser wavelength) versus capillary wave. From the discussion above, it is concluded that several models have been proposed in articles. However, none of the proposed models can account for all the observed LIPSS and their various properties. In fact the LIPSS is a multi parameter mechanism based on surface rippling, acoustic modulation and laser ablation. A complete theory should also explain the variation of periodicity of LSFL and possible change of orientation from HSFL to LSFL, as observed in many experiments.